\documentclass{osa-article}

\journal{osac}
\articletype{Research Article}
\begin{document}

\title{High-precision Quantum Transmitometry of DNA and Methylene-Blue using a Frequency-Entangled Twin-Photon Beam in Type-I SPDC}

\author{Ali Motazedifard,\authormark{1,2,3,4*} and Seyed Ahmad Madani \authormark{1,2,3,5}}

\address{
\authormark{1} Quantum Optics Group, Iranian Center for Quantum Technologies (ICQTs), Tehran, Iran\\
\authormark{2} Quantum Communication Group, Iranian Center for Quantum Technologies (ICQTs), Tehran, Iran\\
\authormark{3} Quantum Sensing and Metrology Group, Iranian Center for Quantum Technologies (ICQTs), Tehran, Iran\\
\authormark{4} Quantum Optics Group, Department of Physics, University of Isfahan, Hezar-Jerib, 81746-73441, Isfahan, Iran\\
\authormark{5} Photonic and Nanocrystal Research Lab (PNRL), Faculty of Electrical and Computer Engineering, University of Tabriz, Tabriz 5166614761, Iran
}

\email{\authormark{*}motazedifard.ali@gmail.com} %% email address is required
%\email{seyed.ahmad.madani@gmail.com}



\begin{abstract}
Using the coincidence-count (CC) measurement of the generated frequency-entangled twin-photons beam (TWB) via the process of type-I spontaneous parametric-down conversion (SPDC) in BBO nonlinear crystal (NLC), we have precisely measured the transmittance of very diluted Rabbit- and Human-DNA, Methylene-Blue (MB), as a disinfectant, and thin-film multilayer at near IR wavelength 810nm with an accuracy in order of $\% 0.01 $ due to the quantum correlation, while accuracy of classical-like measurement, single-count (SC), is in order of $\% 0.1 $ in our setup. 
Moreover, using quantum measurement of the transmittance, the different types of DNA with the same concentration, and also very diluted (in order of pg/$ \mu $l) different concentrations of DNA and MB solutions are distinguished and detected with high-reliability.
Interestingly, in case of Human-DNA samples in contrast to our classical-like measurement we could precisely detect and distinguish two very diluted concentrations $ 0.01\rm ng/\mu l $ and $ 0.1\rm ng/\mu l $ with high reliability while commercial standard spectrometer device of our DNA-manufacturer never could detect and distinguish them. 
Surprisingly, measurement on the thin-film multilayer illustrates that the introduced method in this work might be performed to cancer/brain tissues or Stem cells for cancer therapy, and may hopefully open a pave and platform for non-invasive quantum diagnosis in future. 
\end{abstract}

%\keywords{Spontaneous Parametric Down Conversion (SPDC), Entangled Twin-Photons, DNA, Transmittance, Quantum Diagnosis}

%%%%%%%%%%%%%%%%%%%%%%%%%%  body  %%%%%%%%%%%%%%%%%%%%%%%%%%
\section{Introduction}
%%%%%%%%%%%%%%%%%%%%%%%% P1 :  %%%%%%%%%%%%%%%%%%%%%%%%%%%%%%%%

In the context of the nonlinear light-matter interaction, spontaneous parametric-down conversion (SPDC) is the well-known nonlinear optical process in which a twin-photons beam, a pair of signal-idler photons, is generated out of quantum vacuum when a pump laser beam is incident onto an optical nonlinear material \cite{shySPDC,boyd,agrawal,entanglementReview}. 
In SPDC, an input photon of classical laser pump at higher frequency is annihilated and two entangled photons are created at lower frequencies under the so-called phase-matching (PM) conditions \cite{shySPDC}, i.e., energy-momentum conservation.
SPDC sources in bulk NLCs, as an entangled-squeezed biphoton source (for more details see Ref.~\cite{shySPDC}), are very interested compared to other technologies such as superconductors since they are inexpensive, user-friendly, work at room-temperature, needs no cooling, and finally can be on-chip using integrated quantum photonic techniques in order to be commercialized. 
That is why SPDC in NLCs has been variety range of applications \cite{quantummetrologybook} from the fundamental to the applied point of view such as quantum communication \cite{pirandola2015Communication}, entanglement generation and nonlocal realism tests \cite{epr,nonlocal2,nonlocal3,nonlocal4,aliBellBBO,aliBellPPKTP,nonlocalGenoveseReview2005,polarizationentanglement1995,Forbes}, quantum state teleportation \cite{teleport2001shih,teleportationReview2015}, quantum illumination \cite{qillumination2,qillumination5}, quantum imaging \cite{imagingGenoveseReview2016,mexico1,mexico2,imagingshy}, and quantum interferometry \cite{Mandel1,interferometryGenovese2021} and finally quantum efficiency and absorption measurement \cite{klyshkoEfficiency,absorb1,absorb2,absorb3}.

%%%%%%%%%%%%%%%%%%%%%%%% p3:    %%%%%%%%%%%%%%%%%%%%%%%%%%%%%%%%%%%%%%%%%
Some well-known physical technologies in the context of the quantum light-matter interaction, such as optical spectroscopy, magnetic resonance imaging, and tomography have been employed for understanding biological structure and disease treatment to medical diagnosis. Beyond these, recently, the SPDC-/entanglement-based methods, have attracted interests to be applied in the field of biology and medicine from different points of view such as metrology, phenomenology, philosophy and consciousness \cite{brain1,brain2,brain3,brain4,brain5,brain6,proteinconcentration,human,octbiology}. 
In Refs.~\cite{brain1,brain2,brain3,brain4,brain5,brain6}, the effect of brain tissues with different thicknesses, type and ages on the entangled state has been investigated. High-precision protein concentrometry and tomography of the biological samples such as onion-skin tissue have been done through entangled photons in Refs.~\cite{proteinconcentration} and \cite{octbiology}, respectively. More interestingly, in Ref.~\cite{human} the authors have shown that the Rod cells which respond to individual photons in the human cornea can detect a single-photon incident on with a probability significantly above chance. 
But, in all of them they have not provided any practical optical parameter to characterize the target sample.

%%%%%%%%%%%%%%%%%%%%%%%% p4:  %%%%%%%%%%%%%%%%%%%%%%%%%%%%%%%%%%%%%%%%%
Inspired by the above-mentioned investigations, and using the extension of the Klyshko method \cite{klyshkoEfficiency} we are motivated to measure the transmittance of DNA and Methylene-Blue (MB) solutions through frequency-entanglement, as a practical meaningful optical parameter unlike the Refs.~\cite{brain1,brain2,brain3,brain4,brain5,brain6,proteinconcentration,human,octbiology}.
In cancer therapy and genetic sciences, the measurement of DNA concentrations, identification/detection of the different types of DNAs or even diagnosis of DNAs with different concentrations are essential to biologists and medicines. Furthermore, the MB with different doses as a disinfectant is very important in surgeries. 
Very recently, it has been shown \cite{drug1} a drug developed by Boxin (Beijing) Biotechnology Development LTD based on MB photochemical technology, the so-called BX-1 drug, can inactive lipid-enveloped viruses such as HIV-1 in plasma with high efficiency and without damage to other components in the plasma as a safe and reliable method in clinical trials of HIV treatment. In addition, it has been challengingly confirmed the inactivation effect of BX-1 in SARS-CoV-2 \cite{drug1}. In Ref.~\cite{drug1}, the authors have argued that BX-1 can effectively eliminate SARS-CoV-2 within 2 mins and the virus concentration (titer) decreases and can reach 4.5 log10 $ \rm TCID_{50} $/ml (fifty-percent tissue culture infective dose which is the measure of infectious virus titer). The authors have also argued that BX-1 is safe for blood transfusion and plasma transfusion therapy in recovery patients, and the inactivated vaccine preparation has great potential for treatment in the current Coronavirus outbreak \cite{drug1}. That is one reason we are interested in MB and DNA as a biological and medical/chemical solution samples for its transmittance measurement vs concentration. 
Despite of the quantum spectroscopy, here, the frequency of twin-photons are far from the absorption frequency of the samples which means that it needs no monochromators or very notch filters.
In the following, as an application of quantum measurement in biology we have shown that how a simple method based on the coincidence-count rate measurement of the generated frequency-entangled twin-photons beam due to the type-I SPDC in nonlinear BBO crystal which pass through DNA-and Methylene-solutions enables us to quantum measurement of the transmittance with $ \% 0.01 $ uncertainty \textit{without} any advanced or sub-pixel data analysis. 
\textit{Interestingly}, through the quantum transmittometry Human-/Rabbit-DNA solutions with different concentrations $ \sim (0.01-100) \rm ng/\mu l $ have been detected. 
The DNA solutions with concentrations in order of $ 0.01$ and $0.1  \rm ng/\mu l $ never can be detected by the standard (conventional) commercial spectrometer devices while we could detect them by CC- or transmittance-measurement with more than at lest 10-standardeviation. 
\textit{Surprisingly}, in the same concentration we could distinguish/identify the Human-DNA and Rabbit-DNA solutions via their CC rates which never can be distinguished by classical methods such as single-count measurements. Furthermore, the same procedure has been performed to MB which led to detection of a very diluted Methylene solutions with concentrations $ \sim 4.5 \rm pg/\mu l $ through its quantum transmittometry. 
Finally, the transmittance of some known physical thin-films on soda-lime are measured, that because of similarities to biological tissues, it points out that the introduced method may be applied to cancer-tissues or Stem Cells towards quantum diagnosis.

%%%%%%%%%%%%%%% p5:  %%%%%%%%%%%%%%%%%%%%%%%%%%%%%%%%%%%%%%%%%%%%%%%%%%%%%%%%%%%%
The paper is organized as follows. In Sec.~(\ref{theory}), we review shortly on the physical model of the SPDC and their PM equations. Also, in this section we show that the output of the SPDC is an frequency-entangled state. The experimental setup, methods and results are discussed in Sec.~(\ref{experiment}). 
Finally, we summarize our conclusions and discuss our outlooks in Sec.~(\ref{summary}) .

\section{Theoretical model and description} \label{theory}
However, there is a large texts on the SPDC physics with more details, but, here in this section, we again would like to review short on the SPDC physics to better understanding of the readers especially in the field of biology.

\subsection{Two-particle entangled state}
Entanglement which has no classical counterpart plays the key role in the quantum technologies as the second revolution of the quantum theory during these decades. The best definition which can quantify the entangled state is non-separable property of a quantum state. Let us consider a pure state for a system composed of two spatially separated subsystems 1 and 2 which are closed:
\begin{eqnarray} \label{density1}
&& \hat \rho= \vert \Psi \rangle \langle \Psi \vert , \qquad  \vert \Psi \rangle=\sum_{a,b} \mathcal{C}(a,b) \vert a \rangle \vert b \rangle ,
\end{eqnarray}
where $ \hat \rho $ is the density matrix operator, $ {\vert a \rangle} $ and $ {\vert b \rangle} $ are two sets of orthogonal vectors for subsystems 1 and 2, respectively. 
If $ \mathcal{C}(a,b) $ cannot be factored into a product of the form $ f(a) \times g(b) $, then, the state of the system is non-separable, and thus 
\begin{eqnarray} \label{nonseparable1}
&& \hat \rho \neq \hat \rho_1 \otimes \hat \rho_2 ,
\end{eqnarray}
which means that it is an entangled state \cite{shySPDC}, as was defined for the first time by Schr{\"o}dinger. 
The first simplest example known as the Einstein-Podolsky-Rosen (EPR) state is given by \cite{epr2}
\begin{eqnarray} \label{EPRstate}
&& \vert \Psi \rangle=\sum_{a,b} \delta(a+b-\alpha_0) \vert a \rangle \vert b \rangle,
\end{eqnarray}
where $ a $ and $ b $ are the momentum or the position of particle 1 and 2, respectively, and $ \alpha_0 $ is a constant. 
As can be seen the value of momentum or position is determined for neither single subsystem. Also, the Dirac delta function indicates if one of the subsystems is measured to be at a certain value for that observable the other one is $ \% 100$ determined independent of their distance. 

The \textit{Bell} states (or EPR-Bohm-Bell states), which are the most popular entangled two-particles state, are a set of polarization states for a pair of entangled photons.
They can be represented by
\begin{subequations} \label{Bell1}
	\begin{eqnarray} 
	&&  \vert \phi_{12}^{(\pm)} \rangle=  \frac{1}{\sqrt{2}} [ \vert 0_1 0_2 \rangle \pm \vert 1_2 1_1 \rangle ],\\
	&&  \vert \psi_{12}^{(\pm)} \rangle=  \frac{1}{\sqrt{2}} [ \vert 0_1 1_2 \rangle \pm \vert 1_2 0_1 \rangle ],
	\end{eqnarray}
\end{subequations}
which form a complete orthonormal basis for polarization of light. $ \vert 0 \rangle $ and $  \vert 1 \rangle $ represent the two orthogonal polarization bases, for example, horizontal and vertical linear polarization, respectively. Note that in Eq.~(\ref{Bell1}) the two-photon space-time wavepackets have been ignored. Here, it should be emphasized that to experimentally prepare the Bell states one must overlap these two wavepackets \cite{polarizationentanglement1995,nonlocal4}, i.e., to make them quantum mechanically \textit{indistinguishable}, to keep valid the quantum state \textit{superposition} in Eq.~(\ref{Bell1}).

\subsection{SPDC as a source of two-photon entangled-state}
In nonlinear optical process of SPDC \cite{klyshkobook}, when a coherent pump laser beam is incident onto an optical nonlinear material with second-order nonlinearity, susceptibility $ \bar {\bar \chi}^{(2)} $, a pair of signal-idler photons, the so-called twin-photons, is generated with an efficiency in order of $ 10^{-7}-10^{-11} $. 

Using the first-order perturbation theory together considering the second-order nonlinear interaction Hamiltonian of the SPDC, after some algebraic manipulations the state of the SPDC can be obtained as follows (for more details see Ref.~\cite{shySPDC}) 
\begin{subequations}  \label{spdcstate1}
	\begin{eqnarray} 
	&& \vert \Psi \rangle_{\rm SPDC} \simeq - \frac{i}{\hbar}  \int   dt   \mathcal{\hat H}_{int}    \vert 0 \rangle   \nonumber \\
	&&  = \sum_{s,i} \delta (\omega_p -\omega_s -\omega_i) \Phi(\Delta_k L) \overbrace{ \hat a_s^\dag(\textbf{k}_s) \hat a_i^\dag (\textbf{k}_i) \vert 0 \rangle }^{\vert k_s \rangle \vert k_i \rangle}  \label{spdcstate1a}  \\
	&&  = \! \! \!  \int \! \! \!  g(\omega_p) d\omega_p \!\! \int \!\! d\omega_s f(\omega_p,\omega_s) \overbrace{ \hat a_s^\dagger(\omega_s) \hat a_i^\dagger(\omega_p-\omega_s) \vert 0 \rangle }^{\vert \omega_s \rangle \vert \omega_i \rangle}  \label{spdcstate1b},
	\end{eqnarray}
\end{subequations}
where $ \omega_j $, $ \textbf{k}_j $ ($ j=p,s,i $) are the frequency and wavevector of the signal ($ s $), idler ($ i $), and pump
($ p $) photons, respectively, and also $ \hat a_s^\dag $ ($ \hat a_i^\dag $) is creation operator for signal (idler) photon. Note that $ \omega_p $ and $ \textbf{k}_p $ can be considered as constants. $ g(\omega_p) $ and $ f(\omega_s,\omega_i) $ are the spectral amplitudes of the pump and signal-idler beams, respectively, which usually determined by the wavevector phase-matching condition of the SPDC and provide all information about the spectrum and the correlation properties of the signal-idler pair under the pump photons. 
The sinc-like function $ \Phi(\Delta_k L) $ shows the effect of the finite length of the crystal ($ L $) which is given by \cite{shySPDC}
\begin{eqnarray} \label{sincdunction}
&& L \Phi(\Delta_k L)= \int_0^L dz e^{i\Delta_k L} \frac{e^{i\Delta_k L}-1}{(i\Delta_k L)}, 
\end{eqnarray}
where $ \delta_k= k_p-k_s-k_i $. It is easy to show that in the ideal case, infinite length, $ \Phi(\Delta_k L) \to \delta (\textbf{k}_p-\textbf{k}_s-\textbf{k}_i) $, and thus, the state of the SPDC can be simplified as
\begin{eqnarray} \label{SPDCstatefinal}
&&  \vert \Psi \rangle_{\rm SPDC} \simeq \sum_{s,i} \delta (\omega_p -\omega_s -\omega_i) \delta (\textbf{k}_p-\textbf{k}_s-\textbf{k}_i)  \vert k _s \rangle \vert k_i \rangle. 
\end{eqnarray}
The first Dirac delta in the SPDC state implies on the energy conservation and the latter implies on the momentum conservation. 
More surprisingly, compared to Eq.~(\ref{EPRstate}) the signal-idler two-photon SPDC state of Eq.~(\ref{SPDCstatefinal}) is an EPR type \textit{entangled}-state only because of the delta functions due to phase-matching. It is obvious that it cannot be factorized into a product of the signal and the idler photon state.
By looking at the space-time projection of the state which is equivalent to a Fourier transform, for a
pictorial view of the effective two-photon wavefunction, or wavepacket or \textit{biphoton}, one can emphasize this important physics of this EPR-type-entanglement of the two-photon SPDC-state (for more details see Ref.~\cite{shySPDC}). 

Here, it is worthwhile to remind that the biphoton state which is an EPR-type entangled state is also usually the so-called `\textit{frequency}-entangled' state \cite{shySPDC,quantummetrologybook} (for better understanding see Eqs.~(\ref{spdcstate1a}) and (\ref{spdcstate1b})). Furthermore, one should emphasize that the concept of the biphoton or signal-idler pair photon or twin-photon is never what usually people think, i.e., the two individual independent photons.
\\

\subsection{Phase-matching equations}
The two delta functions in biphoton state of Eq.~(\ref{SPDCstatefinal}) which are the so-called the \textit{phase-matching} (PM) equations determine the energy of the signal-idler pair photons and what they `looks' like and how propagate in space-time. For example, in type-I PM in a negative uniaxial NLC, a linearly polarized pump laser beam as an extraordinary ray under the PM condition can generate a signal-idler pair which both are polarized as the ordinary rays while in the type-II, a signal-idler pair with one ordinary polarization and another extraordinary polarization are generated (for more details see Refs.~(\cite{shySPDC,phasematchBBO,phasematchPPKTP}). The PM equations, the energy-momentum conservation, can be written as
\begin{eqnarray} \label{pm1}
&& \omega_p=\omega_s + \omega_i ,   \qquad    \textbf{k}_p=\textbf{k}_s + \textbf{k}_i . 
\end{eqnarray}
In our case, the planar type-I SPDC in which the signal-idler pair have the same ordinary ($ o $) polarization and pump has extraordinary ($ e $) polarization, the PM conditions can be simplified as follows \cite{shySPDC,phasematchBBO}
\begin{eqnarray} \label{pmfinal}
&& \! k_s(\lambda_s) \! \sqrt{1\! -\left(\frac{\sin\theta_s^{out}}{n_o(\lambda_s)}\right)^2} \! + \! k_i (\lambda_i) \sqrt{1 \!-\left( \frac{\lambda_i}{\lambda_s}\frac{\sin\theta_s^{out}}{n_o(\lambda_i)}\right)^2} \! \! =  \! k_p(\lambda_p,{\rm  \psi}), \nonumber \\
\end{eqnarray}
where $ \lambda_i= \lambda_s \lambda_p / (\lambda_s - \lambda_p) $, $ k_j(\lambda_j)=2\pi n_o (\lambda_j)/\lambda_j $ with $ j=s,i $ and $ k_p(\lambda_p,\rm \psi)= 2\pi n_{\rm eff} (\lambda_p,\rm \psi)/\lambda_p  $. The type-I tuning curves, outside angles versus wavelength, for a NLC can be theoretically plotted using Eq.~(\ref{pmfinal}). Note that the output angles $ \theta_{s,i}^{out} $ are the angles between the pump wavevector $ \textbf{k}_p $ and the signal or idler output beam from the end-side of the crystal. Also, $ \rm \psi $ is the angle between the pump wavevector with the OA inside the crystal. 
It should be noted that the refraction indexes can be calculated from the Sellmeier equations \cite{shySPDC,phasematchBBO,boyd,agrawal} as follows
\begin{eqnarray}
&& n_j(\lambda_j)= \sqrt{L_1 + \frac{L_2}{\lambda_j^2 - L_3} - L_4 \lambda_j^2},  \qquad  (j=e,o)
\end{eqnarray}
where coefficients $ L_m $ ($ m=1,2,3,4 $) are determined experimentally (for example in the case of BBO crystal see Ref.~\cite{bbocoefficient}).
Furthermore, we emphasize that the refractive index for the $ e $-ray also depends on the angle $ \varphi_e = \rm \psi - \theta_s^{int} $ which can be found from Snell's law as $ \sin \theta_s^{int}= \sin \theta_s^{out}/n_o(\lambda_s)$. Therefore, we need to use the effective refractive index for all $ e $-rays (here in our case, \textit{e}-ray is only the pump beam $ p $) which is given by \cite{shySPDC,phasematchBBO,boyd,agrawal}
\begin{eqnarray} \label{n_eff}
&& n_{\rm eff} (\lambda_e, {\rm \psi})= \left[ \frac{\cos^2 \varphi_e}{n_o^2(\lambda_e)} +\frac{\sin^2 \varphi_e}{n_e^2(\lambda_e)} \right]^{-\frac{1}{2}}.
\end{eqnarray}

\subsection{Coincidence count rate}
The joint detection counting rate of detectors $ \rm D_1 $ and $ \rm D_2 $ (for example the first detector is in signal-arm and the other one is in the idler-arm), which is also known as the coincidence count (CC) $ R_c $, on the acquisition time interval $ T $, is the time-integrals over the biphoton wavepacket $ \Psi_{Bp}(r_1,t_1;r_2,t_2) $ as follows \cite{shySPDC}
\begin{eqnarray} \label{r_c1}
&& R_c \! \propto  \! \frac{\eta_1 \eta_2}{\rm T} \! \! \! \int_0^T \! \! \! \!\! \int_0^T  \!\! \! \!  dt_1 dt_2  \vert  \overbrace{ \langle 0 \vert E_2^{(+)}(r_2,t_2) E_1^{(+)}(r_1,t_1) \vert \Psi \rangle_{\rm SPDC }}^{\equiv \Psi_{Bp}(r_1,t_1;r_2,t_2)} \vert^2 ,
\end{eqnarray}
where $ E_j^{(+)}(r_j,t_j)=\int d\omega \hat a(\omega) {\rm exp}[-i(\omega t_j-k(\omega)r_j)]  $ ($ j=1,2 $) is the positive-frequency component of the field operator at detector $ \rm D_j $ with quantum efficiency $ \eta_j $. 
It should be noted that the CC rate can be seen as a criterion for quantum correlation or frequency-entanglement in SPDC systems which reads a counting statistics. Note that in the SPDC-based experiment we usually measure the CC rate to exploit the entanglement. \\

Now, it is worthwhile to illustrate the relation between the CC rate as an experimental quantity and the second-order correlation function $ g^{(2)}(\tau) $, as a photon counting statistics criterion in the theoretical point of view which is given by
\begin{eqnarray} \label{g2}
&& g^{(2)}(\tau)= \frac{\langle \hat a^{\dag}_s(t+\tau) \hat a^{\dag}_i(t) \hat a_i(t) \hat a_s(t+\tau)  \rangle}{\langle \hat a^{\dag}_s(t+\tau) \hat a_s(t+\tau) \rangle  \langle \hat a^{\dag}_i(t) \hat a_i(t) \rangle  }  \quad .
\end{eqnarray}
The correlation function is proportional to the probability of detecting one signal-photon at time $ t + \tau $, given that another idler-photon was detected at earlier time $ t $. $ \tau $ is the delay time of detection between the idler and signal arm in output of the SPDC. 
When $ g^{(2)}(\tau) < g^{(2)}(0) $, the photons tend to distribute themselves preferentially in bunches rather than at random (photon bunching). On the other hand, if $ g^{(2)}(\tau) > g^{(2)}(0) $, fewer photon pairs are detected close together than further apart (photon antibunching).
Using the quantum moment factoring theorem or the Gaussian properties of the noise forces and the quantum regression theorem (see Ref.~\cite{factorizationcite}), the correlation function can be simplified as follows
\begin{eqnarray} \label{g2factor}
&& g^{(2)}(\tau)= 1 + \frac{ R_{si} R_{is}}{R_s R_i} ,
\end{eqnarray}
where $ R_{s(i)}=\langle a^{\dag}_{s(i)}(t) a_{s(i)}(t) \rangle $ is the signal(idler) photon generation rate, and $ R_{si}= R_{is}^\ast=\langle a^{\dag}_{s}(t+\tau) a_{i}(t) \rangle $ is the cross-correlation rates or signal-idler correlation rate.
In Eq.~(\ref{g2}) we have assumed an infinite time resolution of the photon detector and ideal efficiency of the detectors. 
However, in practice in lab, in the detectors, one can only resolve photons in a finite time window $ \tau_w $. 
In the optical frequency, the time resolution is generally within a nanosecond while in microwave is of the order of a microsecond because Raman-absorption and arrival time cannot be distinguished that it sets the length of the detection time window. 
The finite time resolution of the detectors imply that the second-order correlation function is usually a piecewise function as
\begin{eqnarray} \label{g2simple}
&& g^{(2)}(\tau_j)=1 + \frac{\int_{\tau_j}^{\tau_j+\tau_w} \vert R_{si} (\tau)\vert^2 d \tau}{R_s R_i \tau_w}, 
\end{eqnarray}
in which $ \tau_{j+1}=\tau_j +\tau_w $ and can be rewritten as
\begin{eqnarray} \label{g2final}
&&  g^{(2)}(\tau_j):= \frac{R_{cc}(\tau_j)}{R_{acd}}= \frac{R_s R_i \tau_w + \int_{\tau_j}^{\tau_j+\tau_w} \vert R_{si} (\tau)\vert^2 d \tau}{R_s R_i \tau_w} ,
\end{eqnarray}
where the generalized CC rate $ R_{cc}(\tau_j) $ includes two parts: the accidental rate $ R_{acd}=R_s R_i \tau_w $, and the correlated CC rate which contains the contribution of cross-correlation of the signal-idler which is also proportional to $ R_c $ in Eq.~(\ref{r_c1}). 
It should be noted that what we measure in the Lab usually is $ R_c $ in Eq.~(\ref{r_c1}) which can be related to the second-order correlation function or counting statistics through Eq.~(\ref{g2final}).

\section{Experimental results \label{experiment}} 
In this section, we are going to proceed with our experimental results, quantum measurement of the transmittance of the different samples such as DNA, MB and thin-films. Also, we compare the obtained transmittance via the CC-measurement to the SC measurement, as a classical-like measurement, which clarifies the advantage of entanglement-based measurement.

\subsection{Experimental Setup and Adjustments}
\begin{figure}
	\centering
	\includegraphics[width=13cm]{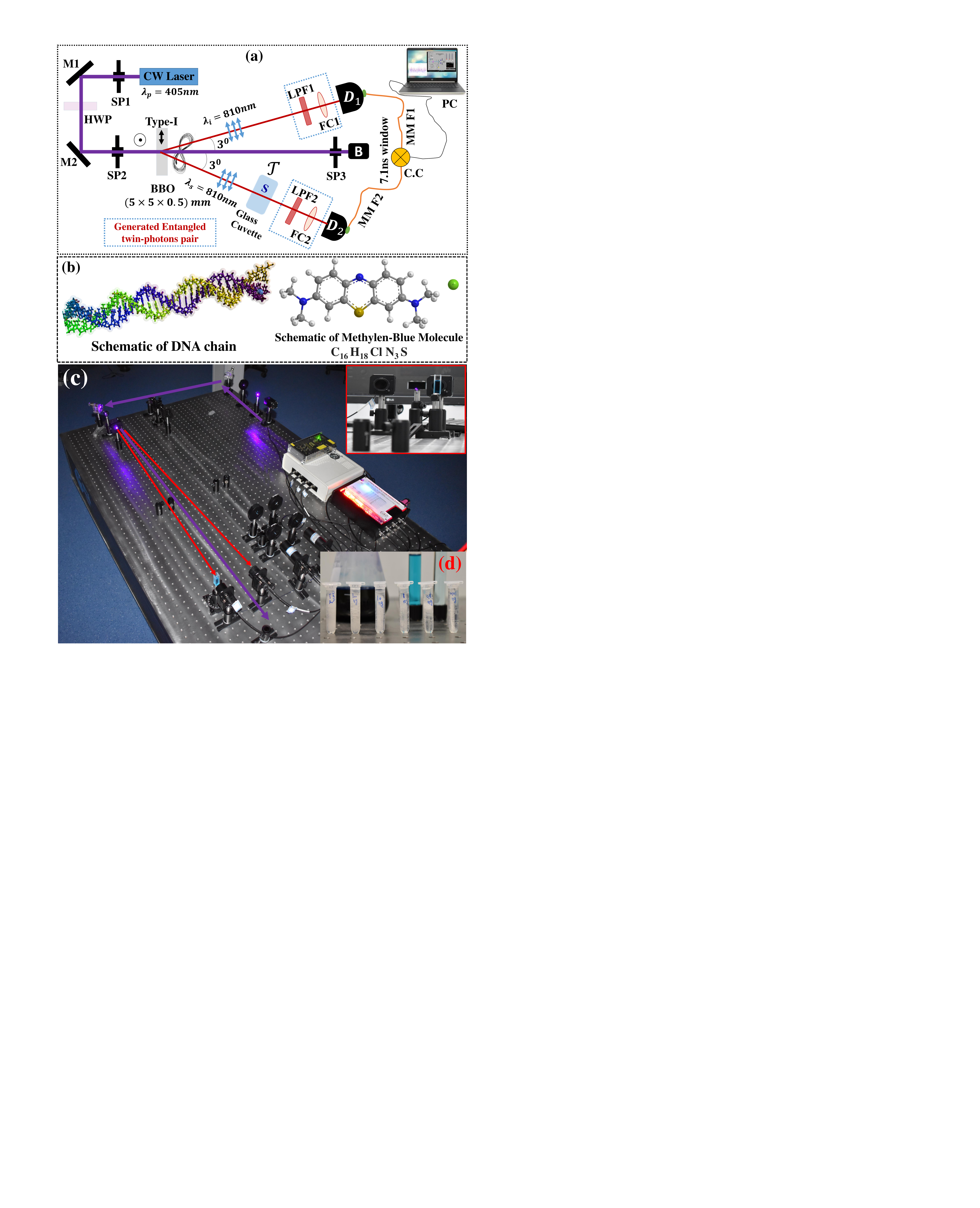}
	\caption{(Color online)  (a) Schematic illustration of the experimental arrangement for quantum measurement of the transmittance of DNA and Methylene-Blue solutions which are inside the glass cuvette and also thin-film samples. (b) Schematic of DNA chain and Methylene- Blue molecule. (c) Top-view of the experimental setup. (d) Different solution samples of Methylene-blue inside Erlen and frozen DNA-solutions with different concentrations inside the micro-tubes at $ -20^{\circ}  {\rm C}$. }
	\label{fig1}	
\end{figure}

Schematic illustration of our experimental arrangement has been shown in Fig.~\ref{fig1}(a). 
A $ 180{\rm mW} $ Gaussian beam of a $ 405{\rm nm} $ diode laser with $ \varnothing 3.8 \rm mm$ after passing through a diaphragm SP1, acting as a spatial filter, passes through a HWP [Newlightphotonic: WPA03-H-810, air-spaced 0th-order waveplate at 810nm, $\varnothing $15.0mm, and OD 1" mounted] to satisfy the PM condition. 
For achieving this, we have set the HWP to $ +3.5^{\circ} $ to have  horizontally e-polarize pump beam. The e-polarized pump beam is reflected from two broadband (380-420nm) high-reflective ($  R>\%99.0 $) mirrors M1 and M2 [Newlightphotonic: BHR10-380-420-45; AOI 45 deg, $ \varnothing $25.4mm] and then passes through the SP2. After SP2, the horizontally \textit{e}-polarized pump beam has power around $ 85{\rm mW} $. 
The pump beam is incident vertically on the single BBO type-I crystal [Newlightphotonic: NCBBO5050-405(I)-HA3] with Size $ (5 \times 5 \times 0.5){\rm mm} $, cut only for type-I PM SPDC and pumped by 405nm diode laser with the half opening angle of 3 degrees, and also AR coated with OD 1" mounted.
Based on the information of the manufacturer company when the pump beam are incident vertically on the entrance-side of the crystal the angle between pump wavevector and the OA is around $ \rm \psi=29.3^\circ $. In this manner, the frequency-entangled signal-idler pair are generated in planer case at output angles around $ \theta_s=\theta_i\simeq 3^\circ $. 
The down-converted photons after passing through an uncoated long-wave pass red filters LPFj ($ j=1,2 $) [Newlightphotonic: LWPF1030-RG715; cutoff wavelength 715nm, $ \varnothing $25mm] by the fiber coupler lenses FCj ($ j=1,2 $) [Thorlabs: PAF2A-11B; FC/APC, f=11.0 mm, 600-1050 nm, $ \varnothing $2.38 mm waist] are focused into the multimode fibers MMFj ($ j=1,2$) [Excelitas: SPCM-QC9FIBER-ND 100 FC], and then transferred via the couplers into the 4-channel single-photon counting module (SPCM) [Excelitas: SPCM-AQ4C; dark counts 351, 483, 156, 146 cps for channels A,B,C, D respectively, time resolution 600ps, dead-time 50ns and quantum efficiency about $ \% $28 at 810nm]. 
To obtain the CC rate of the D1 and D2 we have used the coincidence counter module which we have made it using Altera-DE2 FPGA board and has coincidence time window $ \tau=7.1 $ns with adjustable acquisition time greater than 0.1s. To make sure its validity and also to calibration, we have compared it to the 8-channel coincidence-counter/time-tagger made by QuTools Co. [with time resolution (bin-size) and time-stamp 81ps] such that we found their SCs and CCs were in good agreement. 
After online calculations via the C++ software in our CC module, the CC and SC rates fed to the PC and shown online in the Labview interface panel on the monitor. All the analysis of the experimental data has been done in MATLAB software. Finally, to decrease the background noises we have tried to remove all the reflected and scattered light from the environment and blocked the pump beam after crystal by blocker B. Moreover, we control the temperature of the Lab during each experiment.
Furthermore, to keep laser power constant to avoid noise in count rates, we stabilized the diode laser by a 5V 1A driver with $ \lesssim \% 0.1$ current-fluctuation which results $ \lesssim \% 0.1 $ power-fluctuation by assumption constant load on the adapter. %Thus, during our experiments we are sure that the pump power can be assumed constant.
The distance between two detectors is about 120mm and the distance between detectors and BBO NLC is roughly 98cm.
To calibration and have the optimum PM and maximum CC rate we follow these steps before adding the samples: 

(i) tuning the incident angle and polarization of pump beam to have stable maximum CC rate; 

(ii) adjusting, tilting and moving detectors and BBO crystal which is on tiltable/rotatable mount. Note that the entrance port of the detectors (filter + coupler + entrance edge of fiber are fixed together on a holder as one element) is in the rotatable and movable mechanical elements.

After calibration and achieving the stable maximum CC rates, samples (DNA or MB solutions and thin-films on soda-lime) are added into the setup in the signal-port. The signal beam propagates and transmits through the samples and then reaches to the D2. In this condition, the CC and also SC rates are affected by the transmittance of the samples $ \mathcal{T} $ which depends on the concentration of the solutions or refractive index of the thin-film samples. In the following we show how can measure the transmittance $ \mathcal{T} $ via the CC rate and SC rate as a result of quantum entanglement and classical-like measurements, respectively, to characterize and detect the samples.

\subsection{Samples Preparation}
DNA is the first sample which we are interested in with different concentrations (see Fig.~\ref{fig1}(b)). 
DNA (Deoxyribonucleic acid) is a molecule composed of two chains or strands that are intertwined and wrapped around each other to form a double helix carrying genetic instructions which are responsible for the development, functioning, growth and reproduction of all known organisms and many viruses and plays the key role in genetics. 
These strands are also known as polynucleotides and composed of simpler monomeric units called nucleotides. 
Each nucleotide includes one of four nitrogen-containing nucleobases (cytosine [C], guanine [G], adenine [A] or thymine [T]), a sugar or deoxyribose, and a phosphate group. We choose two different types of DNA, a Human-muscle and a Rabbit-muscle DNA solutions with different concentrations. These samples have been prepared by a PCR (Polymerase Chain Reaction) device. We have prepared Human DNA solutions with concentrations $ 0.01,0.1,1,10 $ and 100 $ { \rm ng/\mu l}$ and Rabbit solution with concentration 100$ { \rm ng/\mu l}$. Prepared samples have been identified via their transmittance using UV-IR Thermo 200C Spectrophotometer device by manufacturer (Neda Chemistry Co.) It should be noted that using the Thermo 200C Spectrophotometer device DNA solutions with concentrations less than 1$ { \rm ng/\mu l}$, we have called it the standard limit (SL) of this device, never can be detected and distinguished. In the other words, our DNA solutions with concentrations 0.01 and 0.1 $ { \rm ng/\mu l}$ never can be detected and identified via their transmittances by the standard commercial spectrometer device Thermo 200C. The prepared samples should be maintained at $-20^\circ \rm C $.

MB [$ \rm C_{16} H_{8} Cl ~\! N_3 S $] with different concentrations is the second type of our sample (see Fig.~\ref{fig1}(b)). Its different doses act as a disinfectant. We prepare the master sample by adding 1.5g MB  in 100ml Ethanol. After that, by adding Ethanol into master Methylene-blue (MMB) one can dilute it and the different concentrations $ 1.5 \rm \mu g/\mu l- 4.5 pg/\mu l$ are prepared to use.

Finally, we prepared some different dielectric thin-films: soda-lime glass without thin-films and also Al and Cr single metal layer and multilayers such as 3-layers $ \rm MgF_2/ZnS/MgF_2 $ and 4-layers $  \rm MgF_2/ZnS/MgF_2/ ZnS $ on the soda-lime glass.  
We prepared Cr/Al nano layers by DC-sputtering coating (by MSS110 DC-Sputtering assisted turbo molecular pump  device made by ACECR in IRAN) on soda-lime glass slides at final vacuum pressure $ 1 \times 10^{-5} $mbar, Ar work-pressure $ 6.6 \times 10^{-3} $mbar, deposition rate $ 3.5 \rm A^{o}/s $, 7SCCM Ar gas, and with current 120mA and voltage 500v at temperature $ 100 ^o C $ around the substrate. 
For the dielectric multilayers we used evaporation method (by EDS110 device assisted diffusion pump made by ACECR) at final vacuum pressure $ 8 \times 10^{-6} $mbar with deposition rate $ 1\rm A^{o}/s $. Here, all thickness have been obtained by the piezo crystal during the coating process with around $\%  10$ accuracy.

In our experiments, DNA and MB solutions are inside the glass cuvette (see inset in Fig.~\ref{fig1}(c)). Also, the thin-films are placed in mount. The cuvette and mount can be fixed on the tiltable and rotatable sample-holder which is placed in the signal-arm in the nearest distance to D2 such that the signal photons incident vertically on the cuvette or thin-film surface and thus transmit without any refraction. To sure it, we couples a 120mW 810nm-diode laser to the fiber coupler FC2 from the back in the signal-port, D2, and try to tilt and rotate the samples to retroreflect incident test-beam 810nm on itself. 

At this point, the setup is ready to measurement of the transmittance (see Fig.~\ref{fig1}(c) which shows the top-view of experimental arrangement and Fig.~\ref{fig1}(d) shows the MMB and diluted MMB inside Erlen and also freezed DNA solutions inside microtubes. The inset shows a diluted MB inside cuvette in front of the D2). In the next subsection, we introduce our simple and short method to measurement of the transmittance.

\subsection{Method}

As mentioned, the samples with transmittance $ \mathcal{T} $ are placed in the signal-arm or sample-arm. 
Interestingly, since the absorption frequency of our samples is roughly far from the signal-idler wavelength 810nm, then in contrast to the quantum spectroscopy, we need no monochromators and narrow-band filters to add in our setup. Therefore, one can assume and sure that all changes in the CC or SC rates is only and mostly due to the transmittance of the samples because of the changes in their refractive index or concentrations. In this manner, using the quantum correlation, i.e., CC rate, we have
\begin{eqnarray} \label{trans1}
&& \eta_2 \mathcal{T}_{cc}= \frac{N_{cc}^{(s)}}{N_1^{(s)}}, 
\end{eqnarray} 
where the quantum efficiency is given by $ \eta_2= N_{cc}^{(0)}/ N_1^{(0)}$ \cite{quantummetrologybook}. $ N_{cc}^{(s)} $ and $ N_{cc}^{(0)} $ are the CC rates in the presence and absence of the samples, respectively, while $ N_1^{(s)} $ and $ N_1^{(0)} $ are the SC rates of the idle-arm (which there is no sample in this arm) in the presence and absence of the sample in sample or signal-arm.
Keeping it in mind that the time average idler SC rate is approximately the same for both cases of the presence and absence of the sample S, i.e., $ \bar N_1^{(s)} \simeq \bar N_1^{(0)}= \bar N_1$. Therefore, the transmittance can be precisely measured using the following relation
\begin{eqnarray} \label{transexact}
&& \mathcal{T}_{cc}= \frac{N_{cc}^{(s)}}{N_1^{(s)}} \frac{N_1^{(0)}}{N_{cc}^{(0)}} ~ ,
\end{eqnarray}
which can be approximated as
\begin{eqnarray} \label{transapp}
&&  \mathcal{T}_{cc} \simeq \frac{N_{cc}^{(s)}}{N_{cc}^{(0)}} ~ . 
\end{eqnarray}
In our experiments we use Eq.~(\ref{transexact}) to obtain the transmittance by calculating the time-dependent CC rate matrix from the experimental data versus acquisition time. To have error in transmittance measurement, we clearly can use the relation 
\begin{eqnarray} \label{errorCC}
&& \frac{\Delta \mathcal{T}_{cc}}{\mathcal{\bar T}_{cc}}= \vert \frac{\Delta N_{cc}^{(s)}}{\bar N_{cc}^{(s)}} \vert + \vert \frac{\Delta N_{cc}^{(0)}}{\bar N_{cc}^{(0)}} \vert + \vert \frac{\Delta N_{1}^{(s)}}{\bar N_{1}^{(s)}} -  \frac{\Delta N_{1}^{(0)}}{\bar N_{1}^{(0)}} \vert  ~ .
\end{eqnarray}
The standard method for transmittance measurement, which is used in the standard commercial spectrometer devices, is measuring the transmitted intensity divide to incident intensity of light, which is fully classical method based on the intensity measurement of the light. Now, for comparison to the classical method for illustration of the advantage of the quantum correlation, frequency-entanglement, respect to the classical intensity measurement one can write a relation for transmittance based on SC rate measurement, which is seen as a best classical-like method, as follows
\begin{eqnarray} \label{transclassic}
&& \mathcal{T}_{sc} = \frac{N_{2}^{(s)}}{N_{2}^{(0)}} ~ , \qquad  \frac{\Delta \mathcal{T}_{sc}}{\mathcal{\bar T}_{sc}}= \vert \frac{\Delta N_{0}^{(s)}}{\bar N_{2}^{(s)}} \vert + \vert \frac{\Delta N_{2}^{(0)}}{\bar N_{2}^{(0)}} \vert , 
\end{eqnarray}  
where $ N_{2}^{(j)} $ (j=s,0) are the SC rates of the signal-arm (sample-arm) in the presence and the absence of the samples. 
Note that if one arm of the SPDC to be blocked, the state of the system to be approximately collapsed to a pure state and thus the other arm can be seen as a non-ideal single-photon source or heralded single-photon source \cite{quantummetrologybook,boyd}. 
Therefore, if by measuring the second-order autocorrelation function via HBT-interferometry (Hanbury Brown and Twiss), one finds $ g^{(2)}(0) <0.1 $ the source is called the idea single-photon source \cite{quantummetrologybook,boyd}, otherwise if $g^{(2)}(0) <1 $ the source is a non-ideal single-photon source.
Therefore, we would stress that in principle $ \mathcal{T}_{sc} $ which is obtained by measuring the SC rate is more accurate than the measurement of the classical light beam intensity to obtain transmittance.

 It is worth to note that the classical analog of the CC measurement of entangled photons is a measurement based on the classically correlated light beam which can be prepared by Arrechi's disk \cite{arrechi} or spatially light modulator (SLM). Surprisingly, it has been shown theoretically and experimentally that the CC-based measurement of the entangled photons is much more accurate than its classical counterpart, CC measurement of classically correlated beam \cite{qillumination2,qillumination5}. 
That is why here we are going to show the advantage and compare the transmittance measurement based on the CC measurement of entangled photons and SC-based measurement.

Finally, it should be noted that we have to take into account the well-known error corrections \cite{errorcorrectionBrida200} due to the accidental coincidence rate, dark count, and dead-time. The accidental rate, $ N_{Ac}=N_1 N_2  \tau_{cc}/T  $, and dark counts $ D_j $ ($ j=1,2 $) can be taken into account \cite{errorcorrectionBrida200} by substituting $ N_{cc}-N_{Ac} $ and $ N_j-D_j $ instead of experimentally recorded counts $ N_{cc} $ and $ N_j $, respectively. Another correction due to the dead time of the detectors can be taken into account by substituting $ \gamma N_j $ instead of $ N_j $ where $ \gamma $ is the dead time correction factor, $ \gamma=1- N_j \tau_{\rm dead}/T $ \cite{errorcorrectionBrida200}. Note that since single count in the sample arm is variable due to the present of the different samples, then its $ \gamma $-factor is not constant and changes in each case but is around 0.9.
Here, $ \tau_{cc} $, $ \tau_{\rm dead} $ and $ T $ 
%(can be tuned between 0.1s to 10s) 
are, respectively, the coincidence time window, the dead time of the detectors and the interval (integration) or gate time which in our case their values are, respectively, 7.1ns, 50ns and 0.3s. Other corrections factors which are discussed in Ref.~\cite{errorcorrectionBrida200} are ignorable in our setup.

%Other correction regarding the dead time, γ, is given by γ≃1-N_SC  τ_dead/T where τ_dead and T are dead time of the detector and interval (integration) or gated time, respectively. Note that since N_SC^((2)) is not constant, then the correction factor γ  in each case changes. In our case, the dead time is about 50ns and the interval time (it can be tuned from 100ms to 10s) was 0.3s, which approximately yields to γ∼0.9, and consequently, we have used γ N_SC^((2))in our calculations. 

\subsection{DNA results}

After calibration of the setup we checked our methods on DNA solutions with different concentrations $ \mathcal{C} $. The results of the transmittance measurement based on CC- or SC-count measurement on Human (H) and Rabbit (R) DNAs are given in Tab.~(\ref{tabDNA}). 
Comparing the four last columns in Tab.~(\ref{tabDNA}) clearly shows that the errors due to transmittance measurement via the CC rate, as a quantum entanglement-based measurement, are very smaller than the SC measurement as a classical-like measurement.  Moreover, relative error ($  \Delta N/\bar N $) in CC rate is generally in order of $ 10^{-4} $ while in SC rate is in order of $ 10^{-3} $, i.e., the CC rates are 1-order more accurate than the SC rates. Furthermore, the average of transmittance obtained by both CC and SC rates is in very good agreement to each other which shows the self consistency of the method.

\begin{table*} %[hbt!]
	\caption{ Experimental values for the transmittance measurement on DNA of Human (H)/Rabbit (R) hand-muscle with different concentrations based on the SC-measurement and CC-measurement. Here, during the experiment the temperature of the Lab is fixed at $ (18 \pm 0.1) ^{\circ} \rm C $. The acquisition time and coincidence time window are respectively $ T_{\rm acq}=0.3\rm s $ and $ \tau=7.1 \rm ns $. The average single-count rate of the idler-arm during the experiment is $ \bar N_1=(1.240 \pm 0.005) \rm Mcps $. Here, the CC and SC rates are vs count per second (cps) or Hz. Also, we define a figure of merit as an entanglement advantage respect to the single-count measurement in count rates as $ \mathcal{G}_{N}={\rm err}^{\rm rel} (N_2)/ {\rm err}^{\rm rel} (N_{cc}) $ and to transmittance measurement as $ \mathcal{G}_{T}={\rm err}^{\rm rel} (\mathcal{T}_{sc})/ {\rm err}^{\rm rel} (\mathcal{T}_{cc})  $ in which the relative error is defined as $ {\rm err^{rel}}(x)=\Delta x/\bar x $.Note that, here, the standard limit of concentration (SL) to detection or identification of DNA through its transmittance is defined as 1$ \rm ng/\mu l $ which is obtained by the standard commercial Thermo200c UV-spectrometer device by manufacturer.}
	%	\begin{ruledtabular}
	\resizebox{\textwidth}{!}{
		\begin{tabular}{cccccccccc}
			\multicolumn{1}{c}{ Sample} & \multicolumn{1}{c}{ DNA }& \multicolumn{1}{c}{Concentration} & \multicolumn{2}{c}{ Count Rates %\footnote{Here, the CC and SC rates are vs count per second (cps) or Hz.}
			} & \multicolumn{2}{c} {Transmittance} & \multicolumn{2}{c}{Ent. Adv. } \\ 
			
			$ \# $ & Type & $\mathcal{C}$ [$ \rm ng/\mu l $] & CC $  N_{cc}$[K cps] & SC $ N_2$[K cps] & $\mathcal{T}_{cc}$[$ \% $] & $\mathcal{T}_{sc}$[$ \% $] & $  \mathcal{G}_{T} $ & $  \mathcal{G}_{N} $ \\ \hline \hline
			
			$1$ & No.samp/cuv. & ----- & $107.17\pm0.05$ & $1005.172\pm0.721$ &  -----  & -----  & ----- & $1.52$ \\ \hline
			
			$2$ & R & $100$ & $90.28\pm0.05$ & $853.118\pm1.001$ &  $84.24\pm0.09$   & $84.89\pm0.16$  & $1.89$ & $1.87$ \\ \hline
			
			$3$ & H & $100$ & $91.50\pm0.03$ & $864.208\pm0.980$ &  $85.37\pm0.07$   & $85.98\pm0.16$  & $3.85$ & $3.84$ \\
			
			$4$ & H & $10$ & $92.53\pm0.01$ & $873.128\pm0.748$ &  $86.34\pm0.05$   & $86.86\pm0.13$  & $2.60$ & $6.43$ \\ 
			
			$5$ & H & $1$ (SL) & $93.28\pm0.01
			$ & $877.985\pm0.920$ &  $87.04\pm0.05$   & $87.35\pm0.15$  & $3.19$ & $13.42$ \\ 
			
			$6$ & H & $0.1$ & $94.16\pm0.03$ & $887.752\pm0.667$ &  $87.86\pm0.07$   & $88.32\pm0.13$  & $1.83$ & $2.24$ \\
			
			$7$ & H & $0.01$ & $94.90\pm0.04$ & $888.613\pm0.774$ &  $88.55\pm0.08$   & $88.40\pm0.14$  & $1.84$ & $2.24$ \\

		\end{tabular}
		%	\end{ruledtabular}
	}
	\label{tabDNA}
\end{table*}

\begin{figure} 
	\centering
	\includegraphics[width=6.50cm]{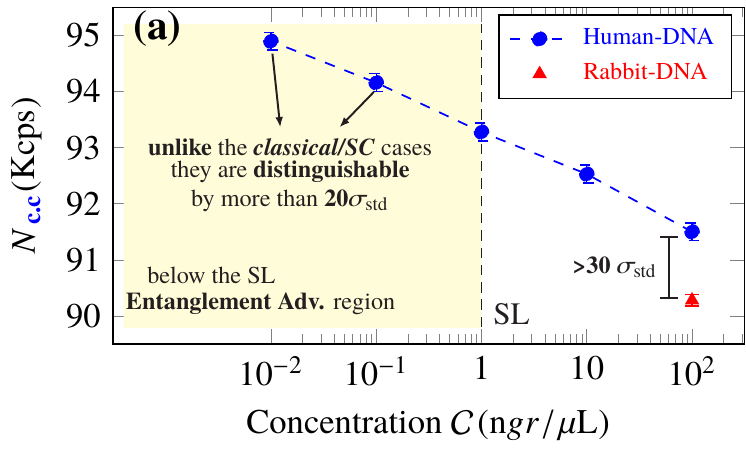}
	\includegraphics[width=6.50cm]{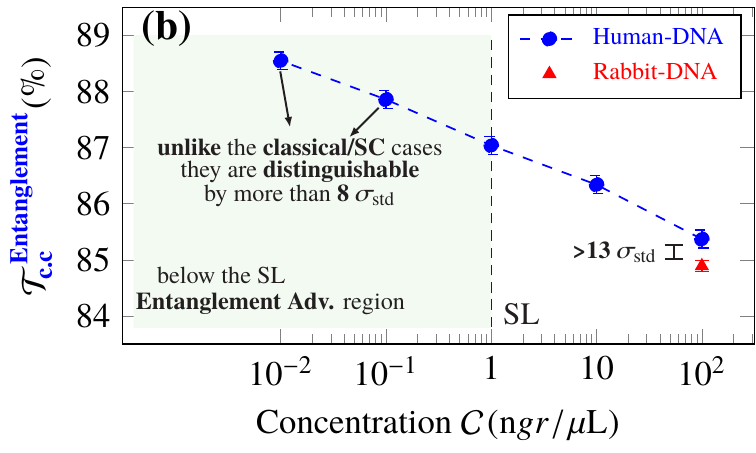}
	\includegraphics[width=6.50cm]{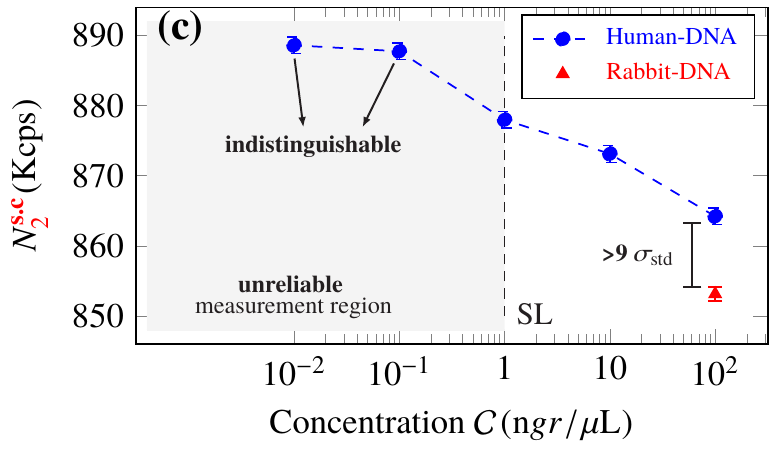}
	\includegraphics[width=6.50cm]{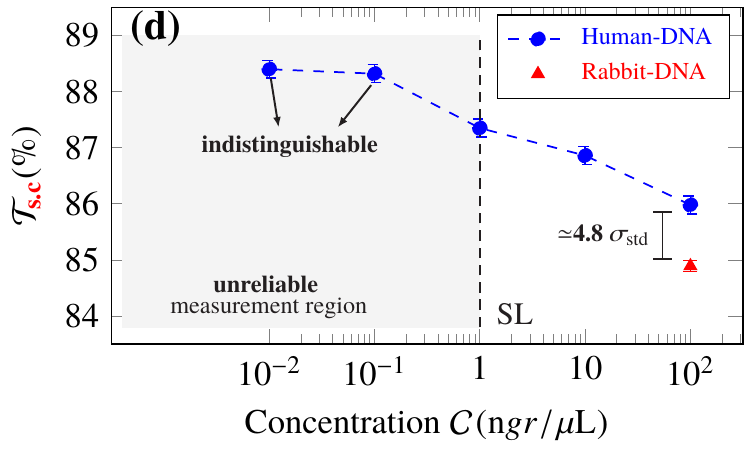}
	\caption{(Color online) (a) and (c) Experimental CC and SC rate measurements for DNA solutions with different concentrations, respectively. (b) and (d) Experimental CC-based and SC-based transmittance measurement of DNA solutions vs their concentrations, respectively. Here, the thick blue circle points and thick red triangle points are respectively referred to experimental data for the Human and Rabbit DNAs. Below the SL region the CC-base measurement results are very reliable in contrast to the SC-based measurement results. }	
	\label{fig2}
\end{figure}

By Comparing the second and the third row in Tab.~(\ref{tabDNA}), one can interestingly figure out the advantage of the use of entanglement or CC rate respect to use of SC rate. As is seen in Fig.~(\ref{fig2}) and Tab.~(\ref{tabDNA}), at the same concentration 100$ \rm ng/\mu l $ the Rabbit-DNA can be detected and distinguished from the Human-DNA by more than 30$ \sigma_{\rm st} $ and 13$ \sigma_{\rm st} $ via the CC rate and its analogous transmittance, respectively, while it can be detected by more than 9$ \sigma_{\rm st} $ and 4.8$ \sigma_{\rm st} $ using SC rate and its analogous transmittance, respectively. 
This shows the entanglement or quantum correlation advantage to high-precision distinguishability between Human and Rabbit DNAs with the same concentration such that the CC-based measurement is approximately 3-times more accurate than the SC-based measurement.  
As is seen in Fig.~\ref{fig2}(c) and (d) and H-rows in Tab.~(\ref{tabDNA}), below the SL of concentration ($ 1\rm ng/\mu l $ which is defined for DNA detection using transmittance measurement via standard spectrometer devices such as Thermo200C) the SC-based measurements never can detect DNA solutions with high reliability, we called it unreliable measurement region. While, surprisingly, using the CC-based measurement (transmittance or CC rate) one can detect and distinguish DNA solutions with high-reliability more than 20$ \sigma_{\rm st} $ and 8$ \sigma_{\rm st} $, respectively, via the CC rate and CC-based transmittance (see Fig.~\ref{fig2}(a) and (b)). In other words, frequency-entanglement enables us to ultra-precise measurement and high-accuracy distinguishability at least 2-order below the SL of DNA-concentration to diagnose and distinguish DNA with concentrations 0.01 and 0.1 $ \rm ng/\mu l $. From the experimental results in Tab.~(\ref{tabDNA}) it is clear that the CC-base measurement of the transmittance has 0.05 accuracy which is at least 3-times more accurate than the SC-based measurement of the transmittance.

By considering the SNR definition as the signal to noise ratio \cite{mexico1,mexico2} and the sensitivity $ \mathcal{S} $ as the minimum noise in the system (see Ref.~(\cite{aliDCEBECforce}) and references therein) as 
\begin{eqnarray} \label{snr}
&& \rm SNR= 10 {\rm Log}(\frac{P_{\rm signal}}{P_{\rm Noise}}) ~ , \qquad   \mathcal{S}=-10 {\rm Log} P_{\rm Noise}^{\rm min}
\end{eqnarray}
where the power of signal and noise is defined as their count rates ($ P_{\rm signal(noise)}:=N_{\rm signal(noise)} $). 
Using the experimental results in Tab.~(\ref{tabDNA}), one can find the average SNR for the CC and SC rates as $\rm  SNR_{\rm Ent.}\simeq 36.5\rm dB $ and $ \rm SNR_{\rm SC}=30.3 \rm dB $, respectively. Moreover, the average sensitivity is respectively $ \mathcal{S_{\rm Ent.}}\simeq -14.2 \rm dB $ and $ \mathcal{S_{\rm SC}} \simeq -29.2 \rm dB $. 
This clearly shows that in our experiment the CC-based measurements which are due to the frequency-entangled state of SPDC has 6dB SNR, and 15dB sensitivity more than the SC-based measurement. Furthermore, in case of transmittance measurement: $ \rm SNR_{\rm Ent.}^{T} \simeq 31.1 \rm dB $ and $ \rm SNR_{\rm SC.}^{T} \simeq  \rm 27.8dB $ which means 3dB more SNR is achievable using CC-based transmittance measurement. As is seen in Fig.~(\ref{fig2}), we should emphasize that different DNAs with different concentrations are reliably detectable and distinguishable from each other using the CC-based measurement with accuracy at least more than 10$ \sigma_{\rm st} $. 

\begin{figure}
	\centering 
	\includegraphics[width=10cm]{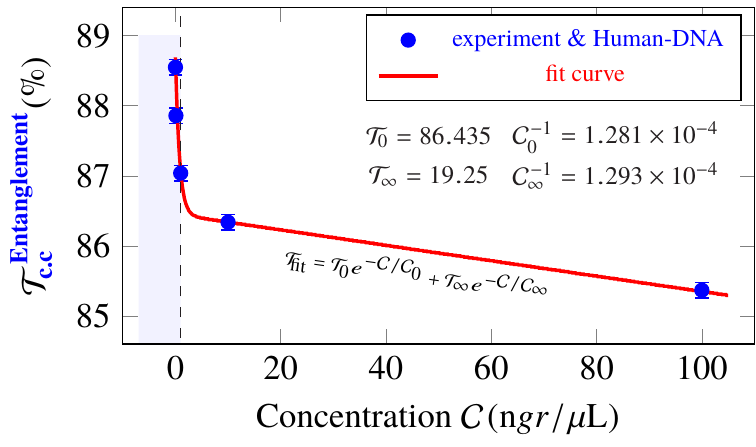}
	\caption{(Color online) Theoretical fitted curve to the experimental transmittance points of DNA obtained from the CC-measurement. }	
	\label{fig3}
\end{figure}

In Fig.~(\ref{fig3}) the theoretical exponential model for transmittance vs concentration as
\begin{eqnarray} \label{t_fit}
&& \mathcal{T}_{\rm fit}= T_0 e^{- \mathcal{C}/\mathcal{C}_0} + T_{\infty} e^{- \mathcal{C}/\mathcal{C}_{\infty}}  ,
\end{eqnarray}
have been fitted to the experimental transmittance points obtained via the CC-measurements. Using this model which has good agreement in this concentration regime, one can estimate concentration of DNA solutions with unknown concentration. This may lead indirectly to high-accuracy DNA concentration measurement, concentrometry. If we assume $ \%  0.01$ uncertainty in experimentally transmittance measurement using the CC rate, one can expect $ 3\rm pg/\mu l $ uncertainty in concentration measurement, while in SC case the uncertainty in transmittance could be $\% 0.1  $ which results  $ 30 \rm pg/\mu l $ uncertainty in concentration measurement using SC rate. Although, we have focused on the transmittance measurement, but using the theoretical fitting curve in Eq.~(\ref{t_fit}), and preparing a database for some DNA solutions in different concentration regimes, one can indirectly measure concentration of DNA solutions via their CC-based transmittance measurement. Furthermore, the concentrometry using CC rate is more accurate than the SC rate notably below the SL which the SC-based measurement cannot work reliably. But, it is clear that the method works better and more accurate for the transmittance measurement instead of concentration measurement.

Finally, this introduced method in this subsection might be used to manufacture an entanglement-based transmittometer/concentrometer simple-device for DNA detection via the CC-measurement.

\subsection{MB results}

\begin{table} 
	\caption{ Experimental values for the transmittance measurement on the Methylene-blue (MB) with different concentrations based on the SC-measurement and CC-measurement. Here, during the experiment the temperature of the Lab is fixed at $ (27 \pm 0.1) ^{\circ} \rm C $. Here, $ \bar N_1=(1.238 \pm 0.005) \rm Mcps $ and the other experimental parameters are the same as Tab.~(\ref{tabDNA}).}
	%	\begin{ruledtabular}
	\resizebox{\textwidth}{!}{
		\begin{tabular}{cccccccccc}
			\multicolumn{1}{c}{ Sample} & \multicolumn{1}{c}{ MB }& \multicolumn{1}{c}{Concentration} & \multicolumn{2}{c}{ Count Rates} & \multicolumn{2}{c} {Transmittance} & \multicolumn{2}{c}{Ent. Adv. } \\ 
			
			$ \# $ & Type & $\mathcal{C}$ [$ \rm ng/\mu l $] & CC $  N_{cc}$[K cps] & SC $ N_2$[K cps] & $\mathcal{T}_{cc}$[$ \% $] & $\mathcal{T}_{sc}$[$ \% $] & $  \mathcal{G}_{T} $ & $  \mathcal{G}_{N} $ \\ \hline \hline
			
			$1$ & No.samp/cuv. & ----- & $59.87\pm0.03$ & $931.34\pm0.33$ &  -----  & -----  & ----- & ----- \\ \hline
			
			$2$ & MMB & $1.5\times 10^4 $ & $2.32\pm0.02$ & $32.13\pm1.23$ & $3.37\pm0.03$ &  $3.44\pm0.14$   & $1.02$  & $1.03$ \\ \hline
			
			$3$ & $ \rm S_0 $ & $4.5\times 10^3$ & $33.23\pm0.01$ & $506.56\pm1.39$ & $55.51\pm0.04$ &  $54.39\pm0.17$ & $4.58$ & $12.54$   \\ \hline
			
			$4$ & $ \rm S_1 $ & $15$ & $51.51\pm0.01$ & $809.26\pm1.84$ &  $86.04\pm0.06$ & $86.89\pm0.23$ & $3.81$ & $9.82$ \\ 
			
			$5$ & $ \rm S_2 $ & $1.5$ & $52.38\pm0.01$ & $820.02\pm1.58$ &  $87.49\pm0.06$   & $88.05\pm0.20$  & $3.37$ & $8.86$ \\ 
			
			$6$ & $ \rm S_3 $ & $0.15$ & $53.27\pm0.01$ & $828.51\pm1.47$ &  $88.97\pm0.06$   & $88.96\pm0.19$  & $2.96$ & $6.80$ \\
			
			$7$ & $ \rm S_4 $ & $0.075$ & $53.89\pm0.01$ & $840.56\pm1.80$ &  $90.00\pm0.06$   & $90.25\pm0.23$  & $3.56$ & $8.83$ \\
			
			$8$ & $ \rm S_5 $ & $0.045$ & $54.58\pm0.01$ & $847.24\pm1.68$ &  $91.17\pm0.06$   & $90.97\pm0.21$  & $3.42$ & $8.81$ \\
			
			$9$ & $ \rm S_6 $ & $4.5\times 10^{-3}$ & $55.42\pm0.01$ & $858.15\pm1.15$ &  $92.56\pm0.06$   & $92.14\pm0.16$  & $2.52$ & $6.22$ \\
			
		\end{tabular}
		%	\end{ruledtabular}
	}
	\label{tabMB}
\end{table}

\begin{figure*} 
	\centering
	\includegraphics[width=6.50cm]{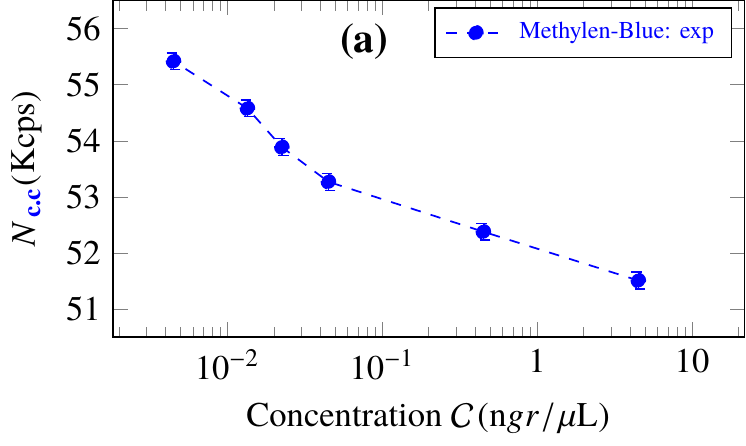}
	\includegraphics[width=6.50cm]{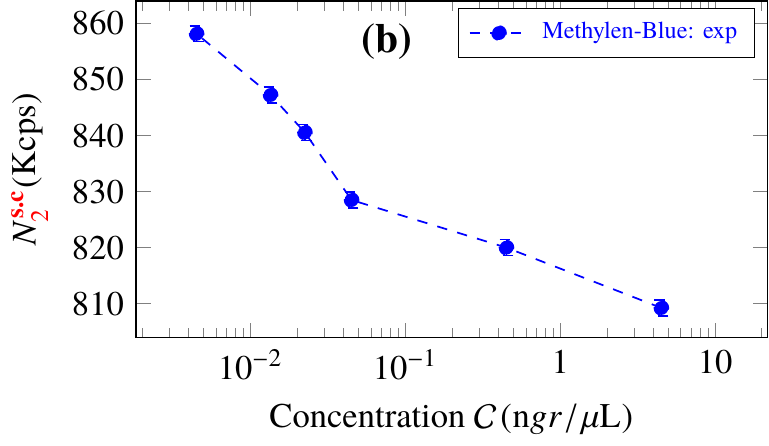}
	\includegraphics[width=6.50cm]{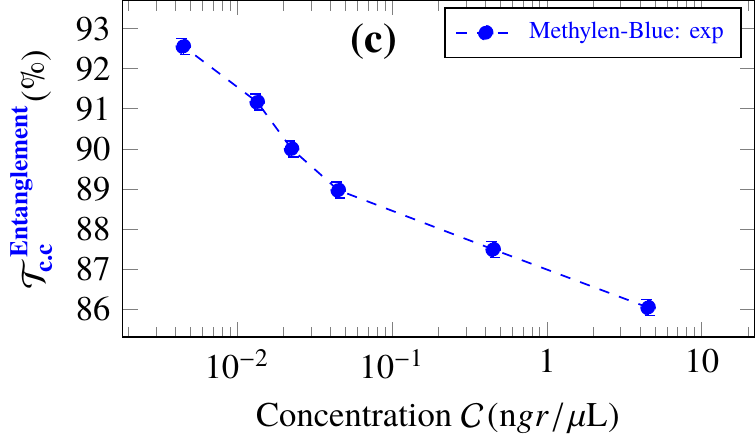}
	\includegraphics[width=6.50cm]{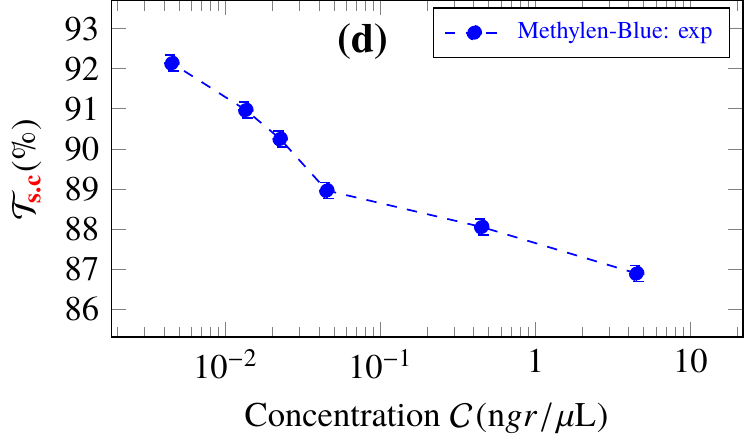}
	\caption{(Color online) (a) and (b) Experimental CC and SC rates for MB solutions with different concentrations, respectively. (c) and (d) Experimental CC- and SC-based transmittance measurement of MB solutions vs their concentrations, respectively. }	
	\label{fig4}
\end{figure*}

Similar to DNA solutions, the same procedure have been performed on the MB solutions to obtain their transmittance vs concentration (see Tab.~(\ref{tabMB}) and Fig.~(\ref{fig4})). As is seen the transmittance obtained in both cases CC- and SC-based measurement are in good agreement. Similar to DNA case the CC-based measurements have more accuracy and more SNR concerning the SC-based measurement. For example, very diluted MB solution with concentration $ 4.5 \rm pg/\mu l $ and $\% 92 $ transmission has been detected through CC-based transmittance measurement with $ \rm SNR\simeq 37dB $ which is 9dB more than SC-based measurement. The physical conclusion in case MB is the same as DNA: the entanglement enables us to more accurate measurement of the transmittance of the diluted solutions with higher SNR and smaller noise. 
The uncertainty in transmittance due to the CC and SC rate measurement is respectively, $ \%0.06 $ and $ \%0.17 $. It means that the CC-based measurement is more accurate and more reliable because of entanglement advantage. Moreover, using the fitting Eq.~(\ref{t_fit}) a $ 0.1 \rm pg/\mu l $ uncertainty is achievable using CC-based transmittance measurement due to the frequency-entanglement advantage.

Here, It should be reminded that as DNA case, one can manufacture an optical device based on frequency-entanglement to high precision measurement of the transmittance or indirect measurement of the concentration through the transmittance. 

Furthermore, as we said in the previous section the maximum absorption wavelength of DNA and MB are about $ \lambda_{\rm DNA}^{\rm ab.}\sim 260\rm nm$ \cite{absorbDNA} and $ \lambda_{\rm MB}^{\rm ab.} \sim 668 \rm nm $, respectively, which are far from signal wavelength $ \lambda_{s}=810 \rm nm $ that passes through the sample. This means that our method for transmission measurement based on the CC rate measurement is still valid.

\subsection{Thin-films results: towards quantum diagnosis of cancer tissues}

To test our method for measuring the transmittance of the transparent bulk object, specially, thin-film (TF) samples, we proceed in this subsection to measure the transmittance of soda-lime glass, single- and multi-layer dielectric thin-films and also single metal layers using CC and SC rate measurement (see Tab.~(\ref{tabTF})). 

To test the validity of our methods, we try to perform measurements on the Soda lime and Lamel glasses (see samples $ \#2 $ and $ \#3 $ in Tab.~(\ref{tabTF})). 
%The refractive index of soda lime glass at wavelength $ \lambda_s= 810\rm nm $ is $\tilde n_{\rm glass}^{\lambda_s} = 1.5170 + \rm i 2.617 \times 10^{-6}  $ with absorption coefficient $ \alpha_{\rm glass}^{\lambda_s}=0.406\rm cm^{-1} $ \cite{refractivecite}. 
Our transmittance results are in agreement with transmittance that we expect for soda-lime with thickness around 1mm and Lamel with thickness around 150$ \mu \rm m $ based on Ref.~\cite{refractivecite}(around $ \% $ 3(2)-4 absorption and $ \% $ 3(2)-4 reflection for soda-lime(Lamel) glass). Moreover, as is seen in Tab.(\ref{tabTF}) both transmittances using CC and SC measurement are in agreement but CC-based measurement has more accuracy than SC-based measurement. 
Using the Fresnel coefficient for transmittance at near-normal incident as a rough model (ignoring the effects of absorption, interference and thickness), $ T=4n_{\rm air} n/(n_{\rm air} + n)^2 $, one can simply find $ n=1.677\pm 0.003 $ and $ n=1.530\pm 0.004  $ using the CC, respectively, for soda-lime and Lamel glasses. It seems in this method one can obtain the refractive index with accuracy in order of $ 10^{-3} $.

Samples $ \#4 $ and $ \#5 $ show the transmittance measurement of dielectric multilayers MgF2/ZnS on soda-lime in which MgF2 and ZnS layers have thickness around $\lambda_{\tiny {1064}}/4 $ with $ 10\% $ tolerance, respectively. 
Their imaginary refractive index are $\tilde n_{\rm MgF2}^{\rm \lambda_s} = 1.4193 + \rm i 4.2103 \times 10^{-4}  $ with absorption coefficient $ \alpha_{\rm MgF2}^{\rm \lambda_s}=65.319\rm cm^{-1} $ and $\tilde n_{\rm ZnS}^{\rm \lambda_s} = 2.3256 + \rm i 0.027886  $ with absorption coefficient $ \alpha_{\rm MgF2}^{\rm \lambda_s}=4326.3\rm cm^{-1} $ \cite{refractivecite} which imply that their absorption wavelength is far from signal-wavelength or they are fully transparent for incident signal-idler wavelength $ 810 \rm nm $. As is seen in Tab.~(\ref{tabTF}), similar to solution cases both SC- and CC-based transmittance measurement are in agreement. However, it is clear that the CC-based measurement is more accurate. 
Here we remind that in cases of glasses and multi-layers samples since the absorption wavelength are far from incident signal-wavelength, thus, our method works well as we expected.

Despite of this to show that our method as we expected cannot work well for absorber material like metals, we try to preform our method on Al and Cr nanolayers with thickness around 15nm and 25nm with $ 10\% $ precision, respectively. Note that both Al and Cr are good absorbers at wavelength 810nm which absorb more than $\% 50 $, especially Al. Refractive index of Al and Cr are, respectively, $\tilde n_{\rm Al}^{\rm \lambda_s} = 2.7670 + \rm i 8.2925  $ with absorption coefficient $ \alpha_{\rm Al}^{\rm \lambda_s}=1.2865 \times 10^6 \rm cm^{-1} $, and $\tilde n_{\rm Cr}^{\rm \lambda_s} = 3.1797 + \rm i 3.4698  $ with absorption coefficient $ \alpha_{\rm Cr}^{\rm \lambda_s}=5.3831 \times 10^5 \rm cm^{-1} $. As is seen in Tab.~(\ref{tabTF}), the results of transmittance based on CC and SC measurement has no agreement which is due to the absorption. In this manner, to have a precise and valid transmittance we need to put narrow-band filters or monochromator to receive exactly the re-emitted photon due to the absorption. That is why the results have no coincidence together. 
%In this manner, the SC-based measurement is more valid than CC measurement.

Surprisingly, it should be noted that our most motivation to perform the method on the TFs is that the similarity between TFs and cancer tissues such as brain tissues which can be prepared as a thin-film on a slide. Therefore, it may be used for cancer therapy or diagnostic purposes. We are hopeful that in the soon future this simple method, which is based on the CC-measurement of generated frequency-entangled biphotons, opens a new pave and simple platform to cancer therapy and quantum diagnosis of cancer more accurate than the present classical methods in different types of cancer tissues.

%As is seen in Tab.~(\ref{tabTF}), in order to show that our CC-based method cannot work around the absorption wavelength of sample, where the absorption wavelength is around the signal wavelength 810nm, we have tried to measure the transmittance of the single Al and Cr layer which their maximum absorption wavelength are $ \lambda_{\rm Al}^{\rm ab.} \sim ??? \rm nm $ and $ \lambda_{\rm Cr}^{\rm ab.} \sim ??? \rm nm $, respectively.

\begin{table*}
	\caption{ Experimental values for the transmittance measurement of TFs. Here, the temperature of the Lab is fixed at $ T=(20 \pm 0.1) ^{\circ} \rm C $, and $ \bar N_1=(1.270 \pm 0.008) \rm MHz $. }
	%	\begin{ruledtabular}
	\resizebox{\textwidth}{!}{
		\begin{tabular}{cccccccccc}
			\multicolumn{1}{c}{} & \multicolumn{1}{c}{ Sample } & \multicolumn{2}{c}{ Count Rates} & \multicolumn{2}{c} {Transmittance} & \multicolumn{2}{c}{Ent. Adv. } \\ 
			
			$ \# $ & Type & CC $  N_{cc}$[K cps] & SC $ N_2$[K cps] & $\mathcal{T}_{cc}$[$ \% $] & $\mathcal{T}_{sc}$[$ \% $] & $  \mathcal{G}_{T} $ & $  \mathcal{G}_{N} $ \\ \hline 
			
			$1$ & No sample & $ 95.50\pm 0.04 $ & $ 933.35\pm 0.74 $ &  -----  & -----  & ----- & ----- \\ 
			$2$ & Soda Lam & $ 89.39\pm 0.01 $ & $ 869.99\pm 0.80$ & $ 93.61 \pm 0.05 $  & $ 93.21 \pm 0.16 $  & 3.50 & 7.01 \\ 
			$3$ & Lamel & $ 91.30\pm 0.02 $ & $ 893.63\pm 0.76$ &  $ 95.61 \pm 0.06 $  & $ 95.74 \pm 0.16 $  & 2.77 & 3.66  \\  \hline
			$4$ & MgF2/ZnS/MgF2/Soda Lam & $ 84.430\pm0.013 $ & $ 821.96\pm 0.49 $ &  $ 88.42 \pm 0.05  $ & $ 88.07 \pm 0.13  $ & 2.72 & 3.96\\ 
			$5$ & MgF2/ZnS/MgF2/ZnS/Soda Lam & $ 65.862\pm0.008 $ & $639.97 \pm 0.82 $ &  $68.97 \pm 0.03  $  & $ 68.57\pm 0.15$ & 4.39 & 10.44 \\ \hline
			$6$ & Al/Soda Lam & $ 16.320\pm 0.002 $  & $ 146.50\pm 0.37 $ & $ 17.09 \pm 0.01 $ & $ 15.70\pm0.05 $ & 7.23 & 25.78 \\ 
			$7$ & Cr/Soda Lam & $ 27.882\pm 0.004 $ & $ 250.51 \pm 0.50 $ &  $ 29.20\pm0.01 $  & $ 26.84 \pm 0.08 $  & 5.48 & 13.60 \\ 
			
		\end{tabular}
		%	\end{ruledtabular}
	}
	\label{tabTF}
\end{table*}

\section{Discussion}
Let us discuss more in this section to illustrate the ability of the introduced simple method.

\begin{itemize}
	\item[(a)] 
	  A considerable difference of coincidence count (CC) rate by human- and rabbit-DNA solutions are actually just due to the difference of their transmittance which itself originates to difference of their refractive indexes at 810nm. It is very important to note that the absorption wavelength of DNAs is usually around 250-300nm \cite{absorbDNA} while our probe wavelength of the signal-idler biphoton is at 810nm which is very far from the absorption frequency of Rabbit and human DNAs.
	 
	\item[(b)] 
	It should be mentioned that the other important physical parameters which may influence or affect on the CC or SC rate are temperature, concentration and inhomogeneity of the solution during the measurement. To avoid the first one, we have fixed the temperature of Lab during the experiment, then, we are sure that any small difference of the CC/SC is just due to the difference of transmittance or small difference of refractive index. 
	Furthermore, to avoid the second one, we requested from our genetic expert colleagues in Neda Shimi Ltd to prepare both human and Rabbit DNA with the same concentration as possible as. 
	Finally, to avoid the effects of inhomogeneity, diffusion, and turbulence of the solutions during the measurement, after pouring solution in the cell, we waited for a certain time to be sure that there is no inhomogeneity, diffusion, and turbulence, and then, we start the data acquisition to obtain the transmittance.
	
	\item[(c)] 
	 It should be noted that the introduced method can be applied to other solutions as far as the absorption wavelength to be far from the signal wavelength. Note that, by having a tunable source like PPKTP or PPLN SPDC sources, which exists in many Labs, one can always apply this method for any solutions or samples. Since it is just enough to tune the SPDC-pump to generate the output wavelength far from the absorption wavelength of the target solution. In this case, in addition to tunable source, one needs to a wide range single photon detectors. In other words, different samples cannot change the physics, and just change the values of the CC/SC rates and consequently the transmittances at desired wavelengths. 
	 
	 \item[(d)] 
	 It should be emphasized that in the present work all transmittance-measurements are a \textit{relative} measurement with \textit{respect} to the prepared concentrations by our reference external molecular genetic Lab, Neda Shimi Co., not an absolute measurement. Furthermore, since our goal is transmittance measurement, then, we have not used or found a theoretical model for behavior of the transmittance vs concentration, temperature, and refractive index of the used samples. But, if we can find a convenient theoretical model similar to Refs.~\cite{concentrationModel1,concentrationModel2} whose authors provided the model versus concentration and temperature for Ethanol, thus, we are able to perform an absolute measurement which yields to absolute values of refractive index or concentration at the fixed temperature. 
	 It would be a good future research plan with collaboration of chemists and geneticists to find a model for the behavior of the DNA and MB to apply the introduced method to obtain the absolute value of the refractive index and concentration.

	 Nevertheless, with no theoretical model, one can still prepare many solution samples with different concentrations at fixed temperature to prepare a database for their transmittance behavior. Then, one can find a trend for each sample versus its concentrations in the fixed temperature. Then, by calibration respect to the used concentration reference at the determined temperature, one can provide relatively concentration and refractive index of an unknown solutions with respect to the reference through their measured transmittance. This may lead to manufacture of a quantum-based device in the future.

	 \item [(e)]
	 Let us clarify the probable ambiguity that why we argue that the thin film multilayer results may lead to quantum diagnosis or application in medicine and biology in the future.

	 In Refs. \cite{brain1,brain2,brain3,brain4,brain5,brain6,proteinconcentration,octbiology,human}, the authors have investigated how much the brain tissue, biological objects, rat or kidney tissues can preserve the frequency-entanglement, polarization-entanglement and even recently the hyper-entanglement such as orbital angular momentum (OAM) which are very interesting from the philosophy and \textit{consciousness} point of view. Their motivations are usually to show that entanglement can be used to characterize these tissues and can be used to cancer therapy or diagnosis of the tissues, or quantum interpret of the brain of biological mechanism. 
	 But, as far as we know, they have not provided a practical measure or quantity for this aim, except Bell parameter S or some well-known nonlocality parameter such entanglement entropies \cite{brain1,brain2,brain3,brain4} which quantify the entanglement. 
	 While we have provided an optical parameter, transmittance, which can practically characterize the sample via its response to the quantum entanglement, nonclassical light. That is why we have been motivated to investigate the thin films samples, since we believe that it may open a new practical way to cancer therapy and quantum diagnosis in medicine and biology.

	 Let us now illustrate the relation or similarities/differences between thin films and cancer tissues or rat sections. As presented in mentioned Refs.~\cite{brain1,brain2,brain3,brain4,brain5,brain6,proteinconcentration,octbiology,human}, the rat sections or brain/kidney tissues with different thickness (10-600$ \mu  $m) are similar to a thin or thick films. But, usually the optical thin films or multilayers are homogeneous and uniform while these biological tissues with different thickness are inhomogeneous and scatterer which can be modeled theoretically by considering different scatterer layers. Although there are this differences between optical multilayers and biological films like brain/kidney/rat tissues, but the thin film model can still provide a rough approximate model to investigate the interaction of entanglement light source with these biological tissues.  
	 That is another reason we have been motivated to perform our method on the thin film layers to measure their transmittance, and we believe that our simple quantum measurement approach can open a new platform to quantum diagnosis of biological tissues and its characterization as has been addressed in Refs.~\cite{brain1,brain2,brain3,brain4}. 
	 Besides these, in Refs.~\cite{brain1,brain2,brain3,brain4} the authors have used the Bell parameter S and entanglement entropy measures which give no information about the materials, while if our method has been applied, then, transmittance measurement enables us to precisely characterize. Furthermore, they have used polarization-entanglement which is very fragile when light passing through the tissues while our frequency-entanglement biphoton is robustness with much higher rate. These are reasons we argue that the thin film results can open a way towards the quantum diagnosis, quantum therapy and also quantum characterization in the context of medicine and biology.
	 
	 \item [(f)] 
	One of the most important questions which can be addressed here is the genuine damage of a biological sample that may be caused due to its transmittance measurement process. 
	Note that we \textit{never} have shown in this paper that a genuine damage is avoided in DNA or MB samples through the use of quantum correlations.  But, it is worth to remind that in SPDC process in NLCs a quantum signal usually has a power in order of pW or smaller which means that it is very low-power compared to the classical light signal which may be in order of $ \mu $W or nW or mW. Thus, one can urge that the damage or destructive effects in case of quantum measurement are definitely very smaller than the case of classical light with very higher power. Therefore, a quantum-based measurement might be seen as a non-invasive measurement technique which is so important especially in biological samples. Nevertheless, it definitely needs to be proven but we leave it as a novel research for the next future works. 
	 
\end{itemize}

\section{Summary, Conclusion remarks and Outlooks \label{summary}}
We have shown that the CC rate measurement of the generated frequency-entangled biphotons at NIR wavelength $ \rm 810nm $ which pass through the solutions, DNA and MB, and TFs enables us to high precision quantum transmittometry with 4-digit accuracy for coincidence time-window 7.1ns, while the SC measurement as the classical measurement leads to 3-digit accuracy. 
Furthermore, for very dilute solutions in order of $ 0.01 \rm ng/\mu l $ or lower, the SC measurement of transmittance never gives reliable answers, while the CC measurement leads to highly reliable measurement of transmittance and distinguishability between the very diluted solutions. 
Also, by obtaining the transmittance, one can calculate the refractive index by considering the proper model for the transmittance vs refractive index and parameters sample yielding to 3-digit accuracy. 
In contrast to the quantum spectroscopy, the method needs no monochromators or narrow-band filters for the samples with absorption frequency far from signal-photon wavelength 810nm. 
Moreover, by choosing a convenient model for concentration vs refractive index or transmittance, one may estimate and obtain the concentration of solutions with $ \rm 0.1pg/\mu l$ precision.

Interestingly, it seems that the method has potential to be improved. If one can use the coincidence counter with more narrower time-window, for example in order of ps instead of current ns, using the deep learning methods for noise cancellation, and using a source with higher rate such as PPKTP Sagnac-based source (see Ref.~\cite{pirandola2015Communication} and references therein) in which $  \Delta N/\bar N $ is usually in order of $ 10^{-5} $, one may expect to measure the transmittance based on the CC measurement with at least 5-digit, which may yield to very more accurate refractive index or concentration.

Finally as outlooks, the method has potential to open a new platform to high-precision quantum transmittometry of different types of samples. Also, we are hoping it opens a new pave in the next future to \textbf{\textit{non-invasive} }quantum diagnosis of cancer tissues like brain tissues or rat, Retina, Liver tissues and Stem Cells as an application of quantum measurement based on entangled photons in the context of quantum biology.

\section*{AUTHOR CONTRIBUTIONS}
A.M.F. proposed the primary idea. A.M.F and S.A.M. developed the idea of using DNA and MB solutions, respectively, and conceived the experiment, built the experimental setup, experimental measurements and analyzed the data. A.M.F. and S.A.M. performed the theoretical calculation and numerical analysis (via MATLAB software), respectively. All authors contributed to prepare the manuscript. A.M.F. wrote the manuscript, revised and answered to the referees' comments during the process.

\section*{Acknowledgments}
We express our gratitude to ICQTs. 
%Moreover, we thank Dr. J. Sabeti, chairman of the board of the ICQTs and Dr. M. Aram, Managing Director of ICQTs, and our colleagues at ICQTs for their helps and supports in realizing this experimental-theoretical work and also for their valuable and constructive comments. 
We also express our gratitude to Dr. M. Abbasi who performed the graphical works related to DNA/Methylene via AVOGADRO software. We also thank Dr. M. Akbari because of his useful comments on quantum spectroscopy. Furthermore, we express our gratitude to Dr. Neda Shirin, manager of Neda Chemistry Ltd., and also to Mr. F. Alidoust and Dr. P. Zardari for preparing DNA and Methylene solution samples, respectively, and their discussions on biological/chemical properties of these solutions. Finally, we thank Prof. A. Rostami, Prof. M. H. Naderi and Prof. R. Roknizadeh because of their useful comments and discussions. We also thank the Dr. N. S. Vayaghan  who is also the supervisor of the quantum optics Lab, and currently the equipment is at his disposal. Finally, we thank the Dr. J. Sabet and Dr. M. Aaram.

\section*{Disclosures}
The authors declare no conflicts of interest.

\end{document}